\documentclass[twocolumn,aps,prb]{revtex4}
\usepackage[T1]{fontenc}
\usepackage[latin1]{inputenc}
\usepackage{amsmath}
\usepackage{amstext}
\usepackage{epsfig}
\usepackage{epic}
\usepackage{eepic}
\usepackage{amssymb}
\usepackage{exscale}
\usepackage{pst-node,array}
\begin{document}
\title{Theoretical studies of the phase transition in the anisotropic 2-D square spin lattice}
\author{Mohamad Al Hajj, Nathalie Guih\'ery, Jean-Paul Malrieu and Peter Wind}
\affiliation{Laboratoire de Physique Quantique, IRSAMC/UMR5626, Universit\'e Paul Sabatier, 118 route de Narbonne, F-31062 Toulouse Cedex 4, FRANCE}
\begin{abstract}
The phase transition occurring in a square 2-D spin lattice governed by an anisotropic Heisenberg Hamiltonian has been studied according to two 
recently proposed methods. The first  one, the Dressed Cluster Method, provides excellent evaluations of the cohesive energy, 
the discontinuity of its derivative around the critical (isotropic) value of the anisotropy parameter confirms the first-order character of the phase     
transition. Nevertheless the method introduces two distinct reference functions (either N\'eel or XY) which may in principle force the 
discontinuity. The Real Space Renormalization Group with Effective Interactions does not reach the same numerical accuracy but it does not 
introduce a reference function and the phase transition appears qualitatively as due to the existence of two domains, with specific fixed points. 
The method confirms the dependence of the spin gap on the anisotropy parameter occurring in the Heisenberg-Ising domain.
\bigskip
\end{abstract}
\maketitle
\section{Introduction}
The study of spin or electron lattices, even when they are governed by simple model Hamiltonians, requires in general approximate methods in order to obtain reliable estimates of the cohesive
energy, of the excitation gap, of the spatial correlation, etc... The treatment of phase transitions is a special challenge for approximate methods since it 
is in general not easy to identify the values of the interactions at the critical points, the nature of the phase transition, as well as the behavior of the properties on both sides of the phase transition. 
The purpose of the present work is to compare the abilities of two methods recently developed by the authors to study a first-order phase transition.\\ 
Despite its rather formal character the spin $\frac{1}{2}$ anisotropic Heisenberg Hamiltonian on an infinite 2-D square lattice
may be used as an excellent model problem to test the ability of a theoretical method to treat a phase-transition phenomenon. This Hamiltonian is given by
\begin{equation}
H= J\sum_{\langle i,j \rangle} (S^{x}_{i}S^{x}_{j}+S^{y}_{i}S^{y}_{j}+\lambda S^{z}_{i}S^{z}_{j}),
\end{equation}
where $\langle i,j \rangle$ runs over all pairs of nearest neighbor sites. This 2-D square lattice model has no exact solution
and has therefore been the subject of numerous calculations \cite{Ref1,Ref2,Ref3,Ref4,Ref5,Ref6,Ref7,Ref8,Ref9,Ref10,Ref11,Ref12,Ref13,Ref14,Ref15,Ref16,Ref17,Ref18,Ref19,Ref20,Ref21,Ref22,Ref23,Ref24,Ref25,Ref26,Ref27} in the recent past, which employ either analytic 
expansions,\cite{Ref13,Ref14,Ref15,Ref16} or numerical algorithms such as Coupled Cluster 
approaches, \cite{Ref8,Ref9,Ref10,Ref11} exact diagonalizations \cite{Ref17,Ref18}
and Quantum Monte Carlo calculations. \cite{Ref19,Ref20,Ref21,Ref22,Ref23,Ref24,Ref25,Ref26,Ref27}
At $\lambda =-1$ a first-order transition takes place between the ferromagnetic phase and a planar-like phase in which the spins in the ground state 
wave function lie in the XY plane. This so-called XY polarized function is such that the sites of one sublattice bear
\begin{equation}
X=(\alpha + \beta )/\sqrt{2},
\end{equation}
where $\alpha $ and $\beta $ are the usual spin up and spin down functions, and those of the other sublattice bear
\begin{equation}
X=(\alpha - \beta )/\sqrt{2}.
\end{equation} 
If one works in the basis of (X,Y) functions instead of $(\alpha , \beta )$ ones, this XY polarized function will appear as the leading configuration for 
the $-1< \lambda < 1$ domain.
At $\lambda =1 $ (isotropic Hamiltonian) a transition to an Ising-like phase occurs. Actually for $\lambda \rightarrow \infty$ the Hamiltonian becomes an Ising
Hamiltonian and the ground state becomes the N\'eel fully spin-alternate function $\Phi _{0}=\alpha \beta \alpha \beta ...$, which is also
the leading configuration for $\lambda >1$.
Early Quantum Monte Carlo (QMC) calculations \cite{Ref6} suggested, although with some imprecision, that this 
transition is of first-order type. More recent and more accurate calculations (see for instance \cite{Ref27}) have confirmed its first-order character. 
One may also quote elaborate Coupled Cluster (CC) calculations \cite{Ref8,Ref9,Ref10} which start from either a planar-like function or the N\'eel 
wave function as reference function $\Phi _{0}$ and assume an exponential form of the wave operator
\begin{equation}
|\Psi \rangle =exp S |\Phi _{0} \rangle,
\end{equation}
where S is restricted to a certain number of local many-body operators (up to 6-body operators).
The results agree very well with those of QMC calculations in the two regions around $\lambda =1$, each region being treated using the relevant reference.
Although the authors do not conclude explicitly, the results support the first-order character of the phase transition at $\lambda=1$.
The extent of the domain of bi-stability is more difficult to assess since it seems to depend on the sophistication of the wave-operator.
The present work studies the same problem using two new methods have different characteristics.			
The methods employed hereafter 
\begin{itemize}
\item[--] the Dressed Cluster Method (DCM \cite{Ref28}) uses, as do the Coupled Cluster expansion (CC) and perturbative approaches, a single reference wave function $\Phi_{0}$, which will be 
either the N\'eel function or the XY polarized configuration. In DCM this 
wave function is used as a bath in which a finite cluster is embedded and treated exactly. Then the configuration interaction matrix relative to the cluster is
dressed under the effect of excitations occurring around the cluster, the amplitudes of which are transfered from the amplitudes of similar excitations 
within the cluster. This approach will be shown to give extremely accurate results, very close to the best Quantum Monte Carlo calculations of the cohesive energy 
and confirm the first-order character of the phase transition but, as well as the CC method, it suffers from the prejudice introduced by the discontinuity of the reference
function $\Phi_{0}$.
\item[--] the Real Space Renormalization Group with Effective Interactions (RSRG-EI \cite{Ref29}) is an improvement of the RSRG method originally proposed by 
Wilson. \cite{Ref30} It proceeds 
through the same reduction of the Hilbert space by considering fragments (or blocks) of the lattice, and a reduction of the Fock space for these blocks to a few
 states of lowest energy. But it extracts effective interactions between the blocks through the exact diagonalization of dimers of blocks. The knowledge of the exact
 spectrum of the dimers enables one to define, using the theory of effective Hamiltonians proposed by 
Bloch, \cite{Ref31} inter-block effective interactions. 
 The method is iterative, it is repeated to blocks of blocks, etc... until it reaches fixed points of the problem. The method provides at a very low cost reasonable
 estimates of the cohesive energy of 1-D or 2-D spin lattices. It does not introduce any reference function, it is therefore in principle continuous on both sides 
 of the critical value of the parameter. However the method leads to two distinct fixed points for the $\lambda <1$ and $\lambda >1$ domains. The 
 iterations result in a discontinuity of the cohesive energy derivative. The method also shows the appearance of an excitation gap for $\lambda >1$.
\end{itemize}
\section{Dressed Cluster Method}
Let us summarize the main points of the Dressed Cluster Method :
\begin{itemize}
\item[--] one first defines a single-determinantal reference function $\Phi _{0}$ on the infinite lattice, namely the N\'eel or the XY function.
For sake of simplicity, the method will be presented here using only the N\'eel function in the $\alpha \beta $ representation
\begin{equation}
\Phi _{0} = \prod_{i} 2i \overline {(2i+1)},
\end{equation}
\item[--] one considers a 2-D square finite cluster of N sites which divides the atoms in two subsets, internal and external, so that the reference
function can be written as
\begin{equation}
\Phi _{0} = \Phi ^{ext}_{0} . \Phi ^{int}_{0},
\end{equation}
\item[--] the model space S is spanned by the determinants obtained from $\Phi _{0}$ by all possible excitation processes
$T^{+}_{i}$ which only concern atoms within the cluster
\begin{equation}
S=\{ \Phi _{i} \} = \{ \Phi ^{ext}_{0}.T^{+}_{i}\Phi ^{int}_{0} \}.
\end{equation}
\end{itemize}
Let $P_{s}$ be the projector onto this model space. The dimension of the full Configuration Interaction (CI) space is equal to that of the isolated cluster. 
Nevertheless the diagonal 
elements of the matrix $P_{s}HP_{s}$ differ from those of the isolated cluster CI matrix under the effect of the embedding, i.e., the energy of each 
determinant is shifted by a quantity $J_{l}$ per alternating bond l, at the frontier.
the determinants $\Phi_{i}$ in the lattice problem interact only with the outer-space determinants $D^{+}_{l}\Phi_{i}$
obtained from $\Phi _{i} $ by a spin exchange $D^{+}_{l}$ on the external bond $l$. Replacing for simplicity the determinants
$\Phi _{i} $ by their index i, the eigenequation for line i is
\begin{equation}
\sum _{j \in S , j \not =i} H_{ij}C_{j} + (H_{ii}-E)C_{i} + \sum_{l \mbox{ ext}} H_{i,D^{+}_{l}i} C_{D^{+}_{l}i }=0.
\end{equation}
The last summation must be evaluated through a proper estimate of the coefficients $C_{D^{+}_{l}i} $. These coefficients are
approximated to the product of the coefficients of the determinants $\Phi _{i}$ by environment-dependent amplitudes $d_{l,i}$
characteristic of the excitations $D^{+}_{l}$ on $\Phi _{i}$.
\begin{equation}
C_{D^{+}_{l}i}=C_{i}.d_{l,i}.
\end{equation}
\begin{figure}[t] 
\centerline{\includegraphics{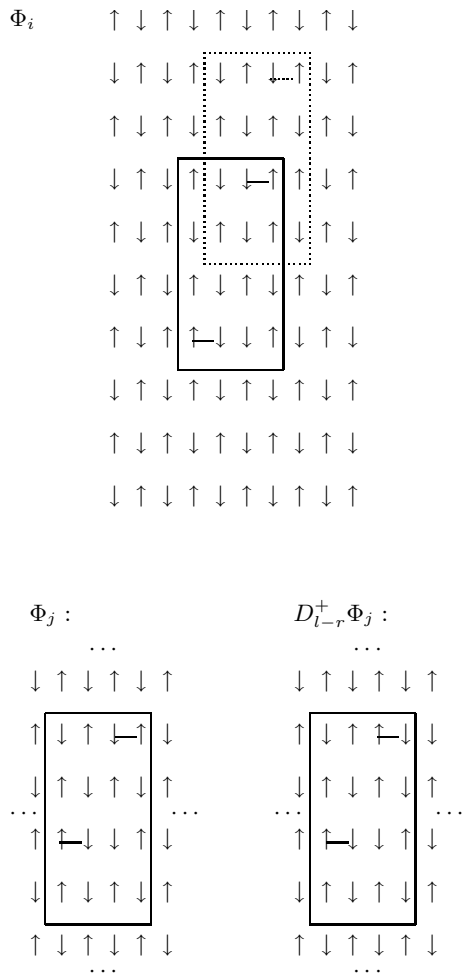}}
\caption{Dressed Cluster Method: schematic view of the principle of the Configuration Interaction dressing under the effect of the spin exchanges
around the cluster, pictured by a full line box. The upper part identifies the cluster and a determinant $\Phi_{i}$, embedded in a N\'eel environment, 
as well as the outer bond l on which a spin exchange will be performed. The lower part pictures the two determinants from which the amplitude 
$d_{l,i}$ (Eq. 10) will be extracted.}
\end{figure}
These amplitudes are extracted from the knowledge of the CI wave function of the embedded cluster.\\
In order to be more explicit, let us consider a determinant $\Phi _{i}$ (cf Fig.1). The cluster is delimited by a continuous-line box and is embedded in
the N\'eel function. Bonds involved in the excitations from $\Phi _{0}$ to $\Phi _{i}$ appear with thick lines. The elementary excitation $D^{+}_{l}$ on an
external bond l (indicated by a dashed line) leads to a determinant $\Phi_{D^{+}_{l}i } $ which interacts with $\Phi _{i}$ through an exchange integral $J_{l}$.
The excitation amplitudes $d_{l,i}$ depend on the environment of the bond l (the largest considered environment is indicated by a dashed-line box) 
and are taken as
\begin{equation}
d_{l,i}=\frac{C_{D^{+}_{l-r}j}}{C_{j}},
\end{equation}
where r is a translation from the external bond $l$ to the outermost equivalent bond $l-r$ of the cluster (which is indicated by a continuous line) 
and $\Phi _{j} \in S $ is such that the environment of the bond $l-r$ in $\Phi _{j}$ has the maximum resemblance with the environment of 
bond $l$ in $\Phi _{i} $.
One must notice that, in some cases, it is necessary to restore the right spin $S_{z}=0$ of the translated determinants
by changing the spins of the atoms furthest from the bond l, in order to obtain the most relevant information from the CI wave function.
Finally the quantity $ \sum_{l \mbox{ ext}} H_{i,D^{+}_{l}i } C_{D^{+}_{l}i } $ can be replaced by
\begin{equation}
(\sum_{l \mbox{ ext}} J_{l}d_{l,i}) C_{i}.
\end{equation}
This summation can be delt with as a diagonal energy shift (dressing)
\begin{equation}
\Delta _{ii}=(\sum_{l \mbox{ ext}} J_{l}d_{l,i}),
\end{equation}
and the corresponding dressing operator $\Delta $
\begin{equation}
\Delta = \sum_{i \in S} \vert \Phi _{i}\rangle \Delta _{ii} \langle \Phi _{i} \vert, 
\end{equation}
Eq. 9 insures the translational invariance; if the determinant $D^{+}_{l}\Phi _{i}$ is identical through a translation $\cal{T}$ to one
of the determinants $\Phi _{k}$ belonging to S i.e., if $D^{+}_{l}\Phi _{i}=\cal{T}$ $\Phi _{k}$, then $C_{D^{+}_{l}\Phi _{i}}=C_{\Phi _{k}}.$
This estimation of $C_{D^{+}_{l}i}$ leads to an important simplification : the effect of excitations on bonds $l$ which are far
from the fragment (by more than the cluster size) is approximated to be identical for all determinants $\Phi _{i} $ and only shifts the diagonal elements of 
the CI matrix by the same amount. It has consequently no effect on the eigenvectors of the dressed CI matrix $P_{s}(H+\Delta ) P_{s}$ and 
can be omitted.
\begin{figure}[t]
\centerline{\includegraphics[scale=0.38]{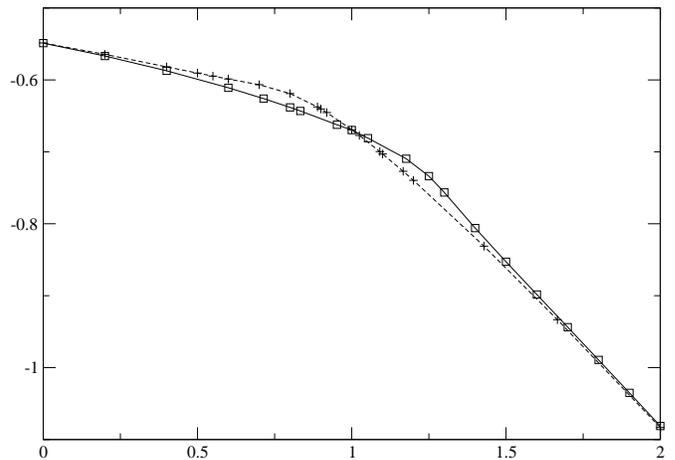}}
\caption{Cohesive energy as a function of the anisotropy parameter $\lambda$. (+) DCM (from a 16-site cluster)
with the N\'eel reference function, ($\Box$) DCM (from a 16-site cluster) with the XY polarized reference function.}
\end{figure}
Since the dressing depends on the eigenvector the procedure must be repeated to self-consistency. One may say that the DC method implies many-body 
operators, up to the number of atoms in the cluster. It does not proceed to a strict exponentialization of the wave function but it employs ratios of
coefficients to transfer information from the internal CI to take into account the effect of elementary excitations on the external
bonds. Through the environmental dependence of these elementary excitation amplitudes, many-body effects are introduced.
The relation with a Coupled Cluster expansion of the wave function \cite{Ref32,Ref33,Ref34,Ref35} has been discussed in ref. 13. The accuracy of
the DC method has been illustrated on 1-D electron and spin (frustrated and non-frustrated) lattices. It has also been applied to the study of
the lowest excitation energies as functions of the bond alternation in the 1-D spin chain. \cite{Ref36}
The DC method is now applied to the 2-D square spin lattice using a $4\times 4$ cluster and starting from both the N\'eel function
and the XY polarized function as reference $\Phi _{0}$. The computed cohesive energy as a function of $\lambda $ is pictured in Fig. 2, where
the two branches, obtained from the XY and N\'eel functions respectively, appear clearly as crossing in $\lambda=1$. One observes the existence
of a continuation of the N\'eel-generated solution in the $0.4<\lambda<1$ and of the XY-generated solution in the $1<\lambda<1.5$. This may be 
seen as the indication of metastable states around the critical $\lambda=1$ value, as expected for a first-order phase transition.\\
The quality of the DCM results has to be assessed by comparison with accurate analytical or numerical calculations. Table I and Fig. 3 
report such comparisons. For $\lambda=1$ our estimate $-0.66928J$ coincides to $10^{-4}$ with the most accurate QMC \cite{Ref25,Ref26,Ref27} value $-0.66944J$. It may be
interesting to compare with CCM results \cite{Ref8,Ref9,Ref10} which are $-0.6670J$ when introducing 6-body operators, and $-0.66817J$ when introducing 
8-body operators.
The difference indicates the importance of many-body operators, and the slow convergence in this expansion. The $3^{rd}$-order spin-wave gives 
$-0.6700J$ and a plaquette expansion \cite{Ref15} $-0.6691J$. \\
The agreement of our DCM values with QMC calculations is similar for $\lambda \not=1$. For $\lambda=0$ we obtain $-0.5489J$, similar to the result 
of Lin et al \cite{Ref27} $-0.54882J$, or for $\lambda=0.6$ (DCM = $-0.61094J$, QMC = $-0.60958J$). Fig. 3 shows the near identity of our results 
with those of Lin et al in the whole $0\leq \lambda \leq 1$ domain. For $\lambda>1$, the agreement in similar, as may be seen from Fig. 3 and 
Table I. For instance we obtain nearly identical values for $\lambda=2$ (DCM = $-1.08329J$, QMC = $-1.08220J$). Notice that we have no convergence 
problem when $\lambda \to 1^{+}$, while cohesive energies could not be obtain in the $1<\lambda<1.09$ domain in ref. 27. 
It is clear that DCM represents, in view of its low cost, an interesting alternative to QMC. 
\begin{figure}[t]
\centerline{\includegraphics[scale=0.38]{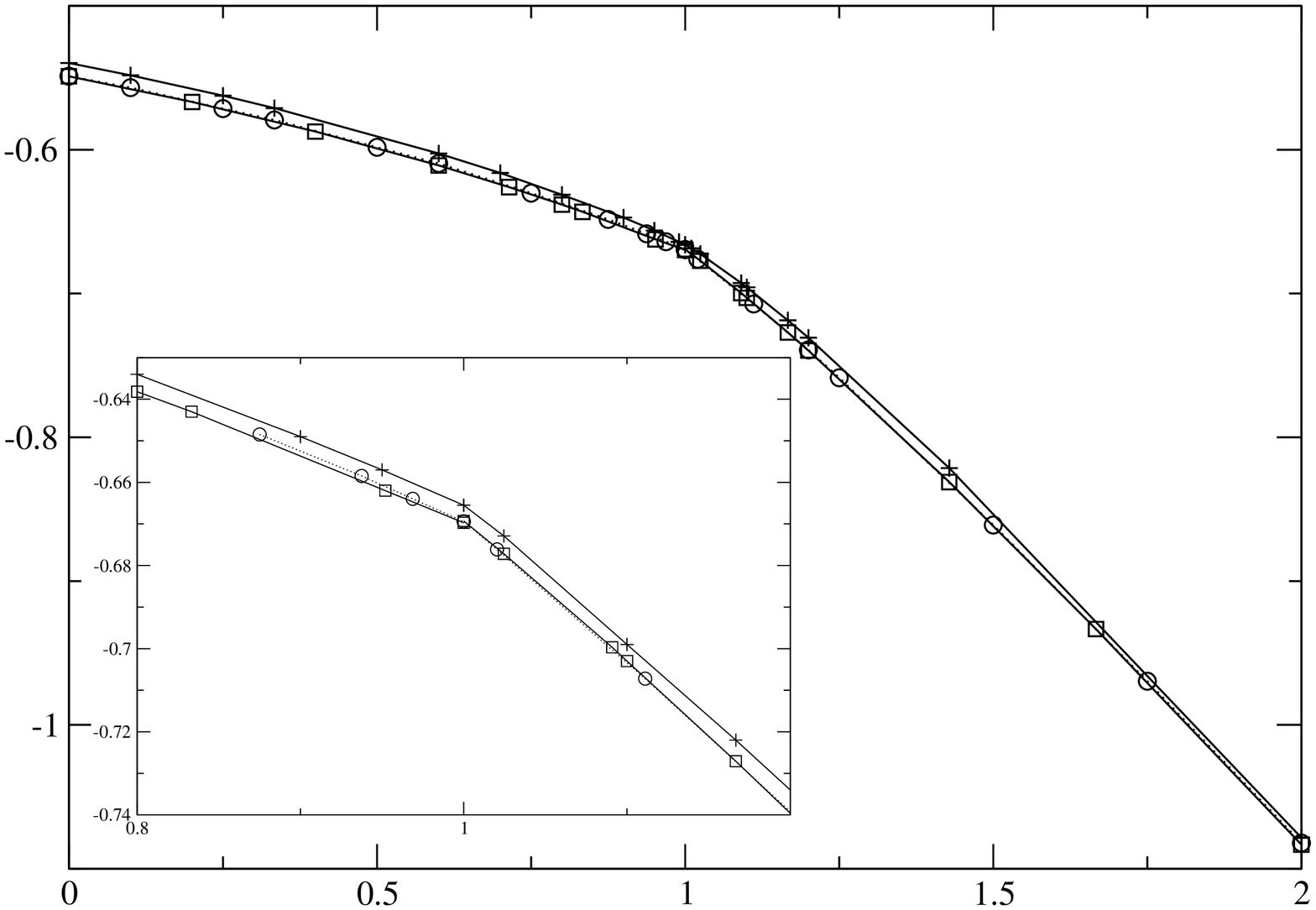}}
\caption{Cohesive energy around the isotropic point. ($\circ$) QMC, \cite{Ref27} ($\Box$) DCM, (+) RSRG }
\end{figure}
\section{Real Space Renormalization Group with Effective Interactions}
\subsection{Method}
The Real Space Renormalization Group proposed by Wilson essentially consists in an iterative truncation of the Hilbert space. The method proceeds
through the definition of blocks of N sites, periodisable fragments of the periodic lattices and the research of the (lowest) eigenstates of the
Hamiltonian relative to these blocks. For a block I, and the corresponding Hamiltonian $H_{I}$ 
\begin{equation}
H_{I}\phi_{K,I}=\mathcal{E}_{K,I}\phi_{K,I}.
\end{equation}
One shall retain a few (let say m) eigenstates of $H_{I}$. Then one will consider a block of blocks (1...I...J...N), and one will approach the 
wave function for this superblock by working in a truncated Hilbert space constituted of all products of the m eigenstates kept for each block.
\begin{equation}
\prod_{I=1,N}\phi_{K,I}, \ K=1,m.
\end{equation}
\begin{table}[t]
\caption{Cohesive energy of the anisotropice 2-D lattice.}
\begin{ruledtabular}
\begin{tabular}{cccc}
  $\lambda$ & DCM & RSRG & QMC \cite{Ref27}  \\
\hline
 0 & -0.54890 & -0.53966 & -0.54882 \\
 0.6 & -0.61094 & -0.60260 & -0.60958 \\
 1 & -0.66972 & -0.66615 & -0.66944  \\
 1.2 & -0.73961 & -0.73072 & -0.73920 \\
 2 & -1.08329 & -1.07849 & -1.08220 \\
\end{tabular}
\end{ruledtabular}
\end{table}
Then the process can be repeated, till convergence. If the blocks and the sets of selected eigenstates are properly defined the problem at each 
iteration may keep its formal structure, while the interactions between the super-super... sites change along the iterations. One then reaches 
in a certain number of steps a fixed point of the problem.\\
This idea is extremely elegant. However the attempts to use it as a practical numerical tool for the study of periodic lattices (of either
spins or electrons) were extremely discouraging. And the method was abandoned, although it gave birth to a deeply different formalism,
namely the Density Matrix Renormalization Group, which is extremely performant, but limited to the treatment of (quasi) 1-D systems. \\
The failure of the RSRG method is due to the simple truncation of the Hilbert space and the total neglect of the non-selected eigenstates of the 
blocks. Rather than trying to treat the effect of the non selected states in a $2^{nd}$-order perturbative 
approach, \cite{Ref37} two of the authors 
have suggested to define effective-interactions between adjacent blocks A and B by solving exactly the Schrodinger equation for the AB dimer, and by making use 
of the Bloch's theory of effective Hamiltonians. We shall not repeat here the formalism, given in ref. 29, which leads to a modified RSRG 
formalism,
called RSRG-EI (EI = Effective Interactions). The first test applications of the method were quite encouraging. We simply make explicit 
hereafter the specification of the method in its simplest version for the study of a square spin lattice.\\
The method consists in considering a square ($3\times3$) block of 9 atoms. Its ground state is a doublet with $S_{z}=\pm1/2$ and it is the only 
state kept hereafter. Let call
\textit{a} and $\bar{a}$ the $S_{z}=1/2$ and $S_{z}=-1/2$ degenerate doublet ground states of the block A. The block can therefore be seen as a super-spin. In order
to establish the effective interactions between the ground states of adjacent blocks, one treats exactly the 18 ($3\times6$)-site superblock AB. One wants to establish 
the effective energies of and interactions between the four products of ground state wave functions which define a model space $ab$, $a\bar{b}$, $\bar{a}b$, $\bar{a}\bar{b}$.
Diagonalizing the exact Hamiltonian for the AB superblock one may identify the eigenstates
$\Psi_{T}^{+}(S_{z}=1)$, $\Psi_{T}^{-}(S_{z}=-1)$, $\Psi_{T}^{0}(S_{z}=0)$, $\Psi_{S}^{+}(S_{z}=0)$
which have the largest projections on the model space,
and their energies $E_{T^{+}}=E_{T^{-}}$, $E_{T^{0}}$ and $E_{S^{0}}$. 
The three energies can be seen as the eigenvalues of a new anisotropic Hamiltonian
\begin{eqnarray}
H_{AB}^{(1)} & = & J_{AB}^{(1)}\lambda^{(1)}(S_{Z_{A}}S_{Z_{B}}-1/4) \nonumber \\
& & +\frac{1}{2} J_{AB}^{(1)}(S_{A}^{+}S_{B}^{-}+S_{A}^{-}S_{B}^{+}) \nonumber \\
& & +E_{A}+E_{B}+\Delta\!E_{AB}.
\end{eqnarray}
Hence 
\begin{equation}
E_{T^{+}}=E_{A}+E_{B}+\Delta\!E_{AB},
\end{equation}
\begin{equation}
E_{T^{0}}=-\frac{1}{2}J^{(1)}\lambda^{(1)}+\frac{1}{2}J^{(1)}+E_{A}+E_{B}+\Delta\!E_{AB},
\end{equation}
\begin{equation}
E_{S^{0}}=-\frac{1}{2}J^{(1)}\lambda^{(1)}-\frac{1}{2}J^{(1)}+E_{A}+E_{B}+\Delta\!E_{AB}.
\end{equation}
From which one obtains
\begin{equation}
J^{(1)}=E_{S^{0}}-E_{T^{0}},
\end{equation}
\begin{equation}
J^{(1)}\lambda^{(1)}=2E_{T^{+}}-E_{T^{0}}-E_{S^{0}}.
\end{equation}
These equations define a new anisotropic Heisenberg Hamiltonian between blocks. The process may be repeated, treating a block of 9 blocks and a superblock of 18 
blocks, till convergence is achieved.
\begin{figure}[t]
\centerline{\includegraphics[scale=0.38]{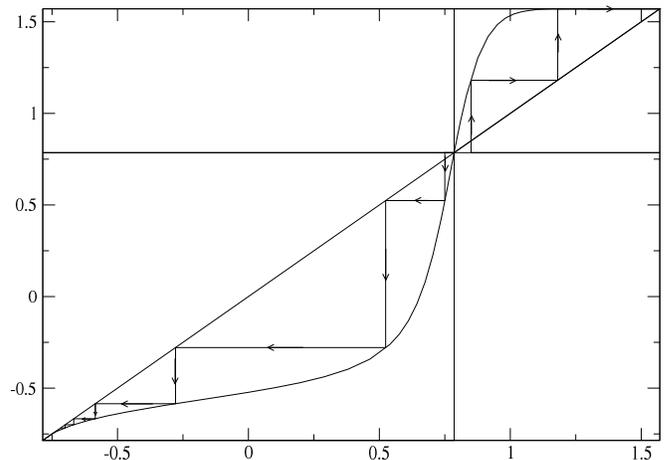}}
\caption{RSRG study: anisotropic parameter $\Phi^{(1)}$ after the first iteration, as a function of the initial anisotropic parameter $\Phi$
(Eq. 20). The stairs illustrate the convergence of the iterative procedure to the fixed points, Ising on the right side, XY on the left side}
\end{figure}
\subsection{Results}
The qualitative key points in that problem are the facts that
\begin{itemize}
\item[(i)] for $\lambda=1$, $\lambda^{(1)}=1$, the problem remains isotropic
\item[(ii)] for $\lambda>1$, $\lambda^{(1)}>\lambda$, the anisotropy is increased in the direction of an Ising problem
\item[(iii)] for $\lambda<1$, $\lambda^{(1)}<\lambda$, the anisotropy increases in the opposite direction towards a pure XY problem.
\end{itemize}
For graphical purposes the anisotropic Hamiltonian may been written as
\begin{equation}
H=J[(S_{z}S_{z})\sin\Phi +(S_{x}S_{x}+S_{y}S_{y})\cos\Phi].
\end{equation}
The isotropic case corresponds to $\Phi=-\pi/4$, the XY problem to $\Phi=\pi/2$, the Ising situation to $\Phi=-\pi/2$. 
On sees that $\lambda=\tan\Phi$. Fig. 4 reports the
evolution of $\Phi^{(1)}$ as a function of $\Phi$. 
The iterative process, starting from as new value $\Phi$ leads to a new anisotropy angle $\Phi_{1}=\Phi^{(1)}(\Phi)$. 
The second step leads to $\Phi_{2}=\Phi^{(1)}(\Phi_{1})$, etc... 
The qualitative nature of the phase transition appears dramatically. Starting from $\Phi>\pi/4$, $\Phi^{(1)}$ increases rapidly.
As seen from Fig. 3 the process converges in a few steps to the 
$\Phi_{n}=\pi/2$ fixed point, i.e, to an Ising problem.
Oppositely, starting from $\Phi< \pi/4$, $\Phi_{1}$ decreases. The fixed point on that side $\lambda < 1$ is the pure XY problem 
($\lambda = -1$, $\Phi=-\pi/4$). But the curve $\Phi^{(1)}=f(\Phi)$ is tangent to the line of slope one $\Phi^{(1)}=\Phi$ for $\Phi=-\pi/2$. 
Hence the fixed point is in principle reached in an infinite number of steps.\\
The quantity $J^{(1)}$ is significantly lower than one for $\lambda < 1$ and tends to zero when $\lambda$ tends to -1. It increases with $\lambda$ 
but remains finite in the region  $\lambda > 1$. Fig. 3 reports the RSRG-EI calculated cohesive energy. For $\lambda = 1$, as already reported 
elsewhere \cite{Ref29} the RSRG-EI cohesive energy is 
$E_{coh}=-0.666155J$. This value is in slightly poorer agreement with the best QMC value $-0.66934J$ than the previously reported DCM value, but it is obtained at 
a much lower cost. The underestimation of the cohesive energy by the RSRG-EI method is systematic but it never exceeds $2\%$ (cf Table I and Fig. 3).
We have carefully checked the existence of a discontinuity of the slopes of the curve $E_{coh}=f(\lambda)$ around $\lambda=1$. 
This discontinuity clearly appears from the insert of Fig. 3. The slope ${(\partial E/\partial \lambda)}_{\lambda \to 1^+}$ between $\lambda=1.02459$ and $\lambda=1$ is $0.32$ in QMC and $0.26$ in RSRG, 
on the $\lambda<1$ side the slope from QMC is $0.175$ (between $\lambda=0.97$ and $\lambda=1$), which the slope from RSRG is
$0.20$ (between $\lambda=0.95$ and $\lambda=1$), $0.21$ (between $\lambda=0.99$ and $\lambda=1$).
Although weaker than the interpolated estimates from QMC, the discontinuity of the slope predicted 
from RSRG-EI is clear.
The existence of a discontinuity was not a priori evident since the
quantities $J^{(1)}$ and $\lambda^{(1)}$ are continuous functions of $\lambda$. The discontinuity comes from the fact that the iterations tend to 
different fixed points for $\lambda>1$ and $\lambda<1$.\\
Actually the method is also able to explain the absence of a gap for $\lambda < 1$ and of the existence of a gap for $\lambda > 1$. For $\lambda < 1$, since one
must repeat an infinite number of iterations with decreasing values of $J^{(1)}$, the lowest states are degenerate. In the $\lambda > 1$ domain, the system will be 
gapped since the process converges in a finite number of steps, with finite values of $J$. Fig. 5 reports the calculated gap for $\lambda$ slightly
larger than 1. We have checked the behaviors of the gap as a function of $\lambda$. Spin-wave theory predicts that it should follow the law
\begin{equation}
\Delta E=2(\lambda^{2}-1)^{1/2}.
\end{equation}
Previous numerical works \cite{Ref21} have shown that the excitation energies are significantly lower, by a factor close to $0.5$. 
Fig. 5 have used an interpolation $0.86634(\lambda^{2}-1)^{1/2}$ which fits well our calculated values.
\begin{figure}[t]
\centerline{\includegraphics[scale=0.38]{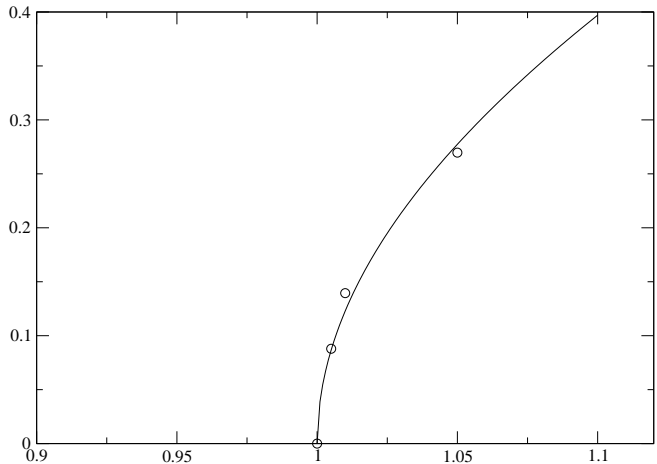}}
\caption{Appearance of the gap in the $\lambda>1$ phase, as calculated from the RSRG-EI. The full line is proportional to $(\lambda^{2}-1)^{1/2}$.}
\end{figure}
\section{Conclusion}
The present paper studies the behavior of a 2-D square spin lattice obeying an anisotropic Heisenberg Hamiltonian. Since it presents a phase transition, this problem
can be seen as a convenient test to compare the abilities of the methods available for the treatment of 2-D (or 3-D) lattices. One may subdivide the methods 
in two groups.
\begin{itemize}
\item[--] methods which rely on (or require the introduction of) a simple zero order wave function. This wave function may be perturbed, or considered as the 
reference function for a Coupled Cluster expansion (i.e an exponential development of the wave operator). In such a case different zero-order or reference 
wave function will be used for the two different phases. This choice of two distinct references may be seen as forcing the phase transition and presents the risk 
to impose artefactual discontinuities. The here employed Dressed Cluster Method only uses the reference function as a bath around a finite cluster, but it is 
subject to the same criticism.
\item[--] prejudiceless methods which do not bias the treatment by introducing reference wave functions. Among them one may quote finite cluster exact 
diagonalization, followed by extrapolations on the cluster size. For 2-D systems extrapolations are quite difficult to perform. Quantum Monte Carlo calculations
require both statistics and extrapolation and the error bars may prevent a clear assessment concerning the nature of the phase transition, when for instance the 
change of the slope of the cohesive energy as a function of the internal parameter is small. Recent progresses have reduced these uncertainties.
\end{itemize}
The excellent agreement of the DCM results with the best QMC calculations for
$\lambda=1$ gives confidence in the accuracy of the calculated dependence of the cohesive energy on the anisotropy parameter and assesses 
the first-order character of the phase transition. \\
The RSRG-EI treatment does not enable one to reach such a numerical accuracy but it presents several advantages
\begin{itemize}
\item[--] it does not introduce the  bias of a reference function
\item[--] it visualizes qualitatively the phase transition in terms of a critical value of the parameter separating two domains with 
their specific fixed points
\item[--] it offers a simple understanding of the gapless-gapped character of the two phases.
\end{itemize}
The philosophy of the RSRG method is responsible for this qualitative and pictorial advantage.
The introduction of effective interactions adds a numerical improvement
to this conceptual tool. Of course, as shown for 1-D lattices, the quantitative performance of the
RSRG-EI treatment is much better when it is possible to
extrapolate its results with respect to the size of the blocks.
This is not possible for the present time for 2-D lattices, since the next size of a square block
would be 25 (which would require the exact treatment of a 50-site problem for the superblock).
But the accuracy of the results from 9-site blocks is surprisingly good and the elegance of the method suggests to consider it as an excellent
exploratory tool.

\end{document}